\documentclass[%
 aapm,
 mph,%
 amsmath,amssymb,
 reprint,%
]{revtex4-2}

\usepackage{graphicx}
\usepackage{dcolumn}
\usepackage{bm}

\usepackage[mathlines]{lineno}

\begin{document}

\preprint{AAPM/123-QED}

\title{Random number generation from a self-chaotic broad-area VCSEL}

\author{Yohann G. Sanvert}
\author{Jules Mercadier}
\author{Stefan Bittner}
\author{Marc Sciamanna}
\affiliation{Université de Lorraine, CentraleSupélec, LMOPS EA-4423 Metz 57070, France}
\affiliation{Chaire Photonique, LMOPS EA-4423 CentraleSupélec, Metz 57070, France}
\date{\today}

\begin{abstract}
The nonlinear dynamics of transverse and polarization modes of a broad-area vertical-cavity surface-emitting laser (BA-VCSEL) exhibit, without any external perturbation, chaos with high correlation dimension, large bandwidth (BW), and good spectral flatness over a wide range of currents. We leverage this for high bit-rate entropy generation and random number generation (RNG), passing the NIST tests with rates up to 150~Gb/s, and observe a correlation between the correlation dimension and the number of passed NIST tests. The RNG shows consistent performance across a wide range of parameters. In contrast to other setups, our system does not require optical feedback or optical injection to generate chaos, making it simple, compact and robust.
\end{abstract}

\maketitle
\indent
Optical chaos from semiconductor lasers has become a valuable resource for generating complex, broadband and highly unpredictable signals. Chaotic dynamics can be induced through various configurations, including external optical feedback \cite{lang1980external,soriano2013complex}, optical injection \cite{kovanis1995instabilities}, or even appear in solitary operation in devices such as single-mode (SM) and multimode VCSELs \cite{virte2013deterministic,mercadier2025chaos,sanvert2025polarization} and broad-area lasers \cite{adachihara1993spatiotemporal} under appropriate conditions. These approaches provide chaos with tunable bandwidth (BW) and flatness. Optical chaotic sources are today exploited in applications ranging from high-speed physical random number generation (RNG) to chaos-based secure communications and cryptography \cite{hirano2009characteristics,tseng2025scalable,virte2014physical,mercadier2026synchronization}. In particular, semiconductor-laser chaos has been widely explored for high-speed physical RNG since the pioneering demonstrations of Uchida $\textit{et al}$. \cite{uchida2008fast} and subsequent works reporting ever-increasing generation rates and improved randomness performances \cite{kim2021massively,li2026solitary,tseng2025scalable}. The combination of high speed, integrability and intrinsic unpredictability makes optical chaos a strong candidate for next-generation information security technologies \cite{Sciamanna2015}.\\
\indent
Among these approaches, VCSELs appear particularly attractive for chaos generation and related applications. Their intrinsic polarization competition can give rise to nonlinear dynamics in free-running operation and has been extensively studied for SM-VCSELs \cite{virte2013deterministic,virte2013bifurcation}. Here, we investigate a BA-VCSEL supporting up to a dozen of transverse modes \cite{bittner2022complex}. So far, only a few studies have addressed the complexity of the chaos generated by a BA-VCSEL compared with SM-VCSELs~\cite{mercadier2025chaos, sanvert2025polarization}.\\
\indent
Our work aims at evaluating the potential of a broad-area VCSEL as a physical random number generator. To this end, we first characterize the dynamical complexity of the laser using several complementary chaos metrics in order to identify operating conditions that maximize complexity and favor RNG operation. Such a systematic analysis is largely absent from previous BA-VCSEL studies \cite{lu2024parallel}. Also, compared to SM-VCSELs, BA-VCSELs exhibit a broader parameter range with chaotic dynamics, naturally offering greater diversity for RNG implementations \cite{sanvert2025polarization}. Moreover, unlike many photonic RNG schemes requiring complex external configurations, our device operates in a free-running solitary regime using an off-the-shelf laser, eliminating time-delay signatures typical of feedback systems and enabling a simpler and more robust architecture suitable for practical implementations.
\section*{Experimental chaos characterization}
\begin{figure}[t]
	\centering
	\fbox{\includegraphics[width=\linewidth]{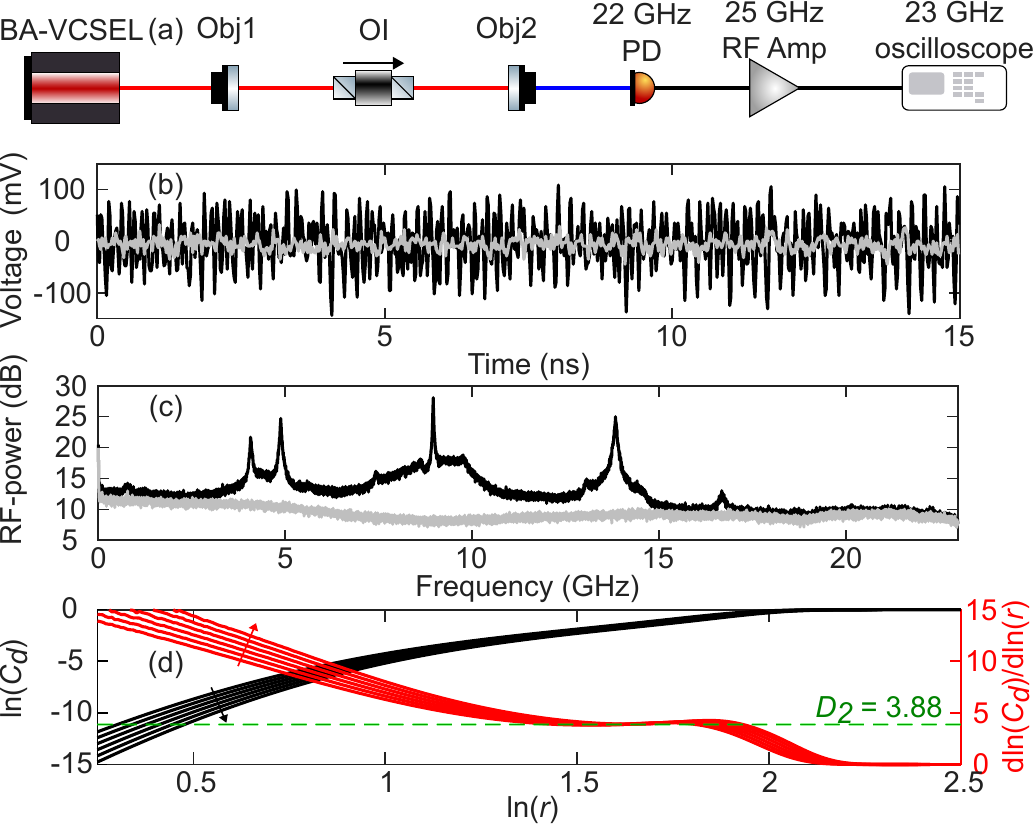}}
	\caption{(a) Experimental setup. Obj1: $40\times$ objective (NA = 0.6), OI: optical isolator, Obj2: $20\times$ objective (NA = 0.5), PD: Photodetector, RF Amp: RF amplifier. (b) Time trace and (c) RF-spectrum of the laser signal at 4.89 mA (black) and instrument noise (gray). A moving average of 5~MHz width is applied to the RF-spectrum. (d) Results of the Grassberger-Procaccia algorithm for an injection current of 4.89~mA with embedding dimensions increaing from 25 to 30 as indicated by the red and black arrows, and an embedding delay of 0.1~ns. The correlation integral (its logarithmic derivative) is plotted in black (red) with respect to the sphere radius $r$.}
	\label{fig:fig1}
\end{figure}

\begin{figure}[t]
	\centering
	\fbox{\includegraphics[width=\linewidth]{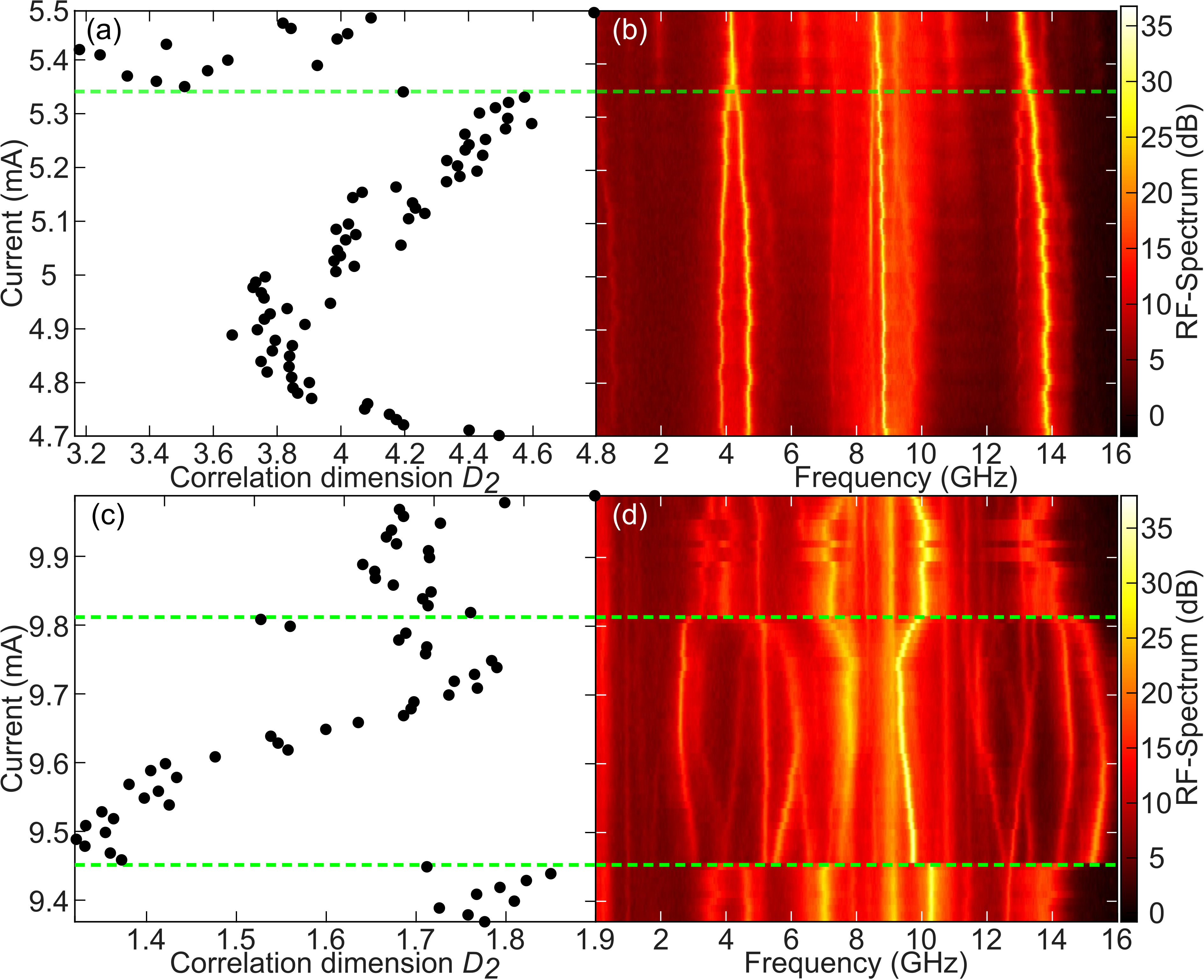}}
	\caption{Comparison between the correlation dimension and the RF-spectrum for (a,b) $I\in$[4.7,5.5]~mA and (c,d) $I\in$[9.37,9.99]~mA. For all currents, the parameters for the GP algorithm are chosen as explained in Appendix \ref{GPAlgo}. The green doted lines indicate sudden changes in the RF-spectrum and in the evolution of the correlation dimension.}
	\label{fig:fig2}
\end{figure}

\indent
Figure~\ref{fig:fig1}(a) shows the setup similar to the one in Ref.~\cite{mercadier2025chaos}. The system under study is a BA-VCSEL with a circular aperture of $15~\mu$m diameter (Frankfurt Laser Company FL85-F1P1N-AC) pumped through a ring contact with a current threshold of 1.8 mA. The laser beam is collimated by a $40\times$ microscope objective. An optical isolator prevents back-reflections into the VCSEL and allows to select the polarization. Our VCSEL emits along two orthogonal polarization axes. The isolator’s polarizer is oriented at $45^\circ$ relative to these axes to measure a superposition of both polarization states. Lastly, the laser beam is coupled with a $20\times$ objective (Obj2) into a multimode fiber (Thorlabs M116L02)  that is connected to a photodetector (Newport 1484-A-50, 22~GHz), a RF amplifier (SHF S126A, 25 GHz, 29~dB) and an oscilloscope (Tektronix DPO72340SX—23~GHz, 50~GS/s).\\
\indent
Figures~\ref{fig:fig1}(b) and \ref{fig:fig1}(c) show the time trace and the RF-spectrum of the output power at an injection current of 4.89~mA. We use the following metrics to characterize the RF-spectra in the following: the traditional chaos bandwidth (BW) is defined as the maximum frequency $f_{m}$ for which 80\% of the total RF power is contained between 0 and $f_{m}$. In contrast, the effective BW is the frequency interval that includes the most prominent spectral components whose cumulative power reaches 80\% of the total RF power \cite{lin2012effective}. We also calculate the spectral flatness, defined as the ratio of the geometric mean to the arithmetic mean of the RF spectrum components \cite{johnston2002transform}. For this specific current, the laser dynamics is chaotic with a traditional chaos BW of 13.8~GHz, an effective chaos BW of 2.8~GHz and a flatness of 0.29. These values are modest compared to those achieved with chaos from optical feedback \cite{bouchez2021optimized} or optical injection \cite{mercadier2024optical}, but are high for a chaos achieved from bifurcations involving only intrinsic laser time scales, which are the birefringence frequency and the relaxation oscillation frequency in our case \cite{sanvert2025polarization,mercadier2025chaos}.\\
\indent
Next, we use the Grassberger-Procaccia (GP) algorithm to estimate the fractal dimension for $I = 4.89$~mA \cite{grassberger1983measuring}. The results for embedding dimensions varying from 25 to 30 and an embedding delay of 0.1~ns are shown in Fig.~\ref{fig:fig1}(d), yielding a correlation dimension $D_{2} \simeq 3.88$, used here to estimate the fractal dimension. To choose the correct value for the embedding delay, we use the first minimum of the computed mutual information. We additionally apply a Theiler window of 5~ns since it removes short time correlations, providing a better estimation of the fractal dimension \cite{theiler1986spurious}. Lastly, we applied singular spectrum analysis to reduce the measurement noise in our time traces \cite{golyandina2001analysis} (see Appendices \ref{AppendixNoiseTitration} and \ref{GPAlgo}).\\
\indent
Figure~\ref{fig:fig2} shows the evolution of the correlation dimension for two current ranges featuring chaotic dynamics, 4.7 to 5.5~mA and 9.37 to 9.99~mA. For both current regimes, the corresponding RF-spectra are also plotted. First, one observes that in both current ranges, the correlation dimension varies significantly, ranging from 3.2 to 4.8 in the first current region and from 1.3 to 1.9 in the second. Secondly, it is very interesting that by only changing the injection current, the correlation dimension can vary by more than a factor of two between different chaotic zones. Figure~\ref{fig:fig2} also reveals that the correlation dimension is strongly linked to the dynamics observed in the RF-spectrum. Indeed, around 5.35~mA, an abrupt transition in the correlation dimension is observed when the two frequency components of the chaotic dynamics around 4~GHz merge. At this current, the correlation dimension transitions from a smooth evolution in a U-shape to an irregular evolution. Similar transitions occur at 9.45 and 9.81~mA: a clear change of the frequency components in the RF-spectrum coincides with an abrupt transition in the correlation dimension.\\
\indent
This analysis highlights the richness of the dynamics in a BA-VCSEL. The nonlinear dynamics of transverse and polarization modes in a BA-VCSEL results in additional polarization switching points compared to SM-VCSELs, each of them leading to new bifurcations \cite{bittner2022complex}. We emphasize the advantages of BA-VCSELs compared to SM-VCSELs as the former exhibit more distinct and larger chaotic zones when the injection current is varied \cite{sanvert2025polarization,mercadier2025chaos,bittner2022complex,virte2013bifurcation}. Thus, BA-VCSELs can generate several chaotic regions with different dynamics with very different correlation dimensions which may be exploited for diverse applications such as cryptography \cite{mercadier2026synchronization} or RNG. \\
\indent
Next, we study the BW \cite{lin2012effective}, spectral flatness \cite{johnston2002transform} and entropy generation \cite{hart2017recommendations,kim2021massively} of our BA-VCSEL for the same current ranges as before. Figure \ref{fig:fig3} shows the traditional and effective BW, the spectral flatness and the entropy generation for injection currents between 4.7 and 5.5~mA as well as between 9.37 and 9.99~mA. These studies are motivated by applications of our VCSEL for RNG since higher BW and spectral flatness often lead to a higher entropy rate which is required for high-speed RNG \cite{tseng2025scalable,hart2017recommendations}. In our case, we observe that the BW and the spectral flatness vary both inside and between the chaotic regions, highlighting the evolution of the dynamics with the current. We note that the effective BW is clearly smaller than the traditional BW, which is expected given the definition of these two BWs and considering that the RF-spectra exhibit strong frequency components at the birefringence frequency and its harmonics. However, the traditional BW indicates that there are nonetheless relevant frequency components at higher frequencies. \\
\indent
To estimate the entropy generation shown in Fig.~\ref{fig:fig3}(c), the Cohen-Procaccia algorithm is used. This algorithm computes the estimated entropy rate $h_{\text{CP}}$ as a function of the number of digits $N_{d}$ and the sampling rate $\tau$ as explained in \cite{kim2021massively,hart2017recommendations} and in Appendix \ref{EGEstimation}. For Fig.~\ref{fig:fig3}(c), the entropy generation for $N_{d} = 6$ is shown as function of the injection current. In the whole current range, $h_{CP}$ varies between 144 and 158~Gb/s. This indicates that, with proper post-processing, all injection currents in the two chaotic regions may lead to fast RNG around 140~Gb/s. It is worth noticing that while the BW varies by a few GHz, the spectral flatness by a few percent, and the correlation dimension by more than a factor of two, the entropy generation rate in Fig.~\ref{fig:fig3}(c) varies by only around 10~Gb/s in the full analyzed current range, which is fairly little.
\section*{Random Number Generation}
\indent
Finally, as an application of the chaotic dynamics of a BA-VCSEL, we investigate its performance as a random number generator. The sampling rate is set at 1~Gb/s and each sample is converted into a binary value by thresholding with the mean value. To improve the statistical balance of the resulting bit stream, a 40~ns delayed XOR operation is applied. This rather basic post-processing allows us to focus on the general trends as function of the pump current more than specific cases for which post-processing will be refined. \\
\indent	
Firstly, we analyze the relation between the correlation dimension and the percentage of passed NIST SP 800-22 tests \cite{sonmez2016recommendation}. Figure~\ref{fig:fig4} plots the percentage of passed NIST tests with respect to the correlation dimension for currents between 4.7 and 5.34~mA and between 9.37 and 9.99~mA. We observe a correlation between the percentage of passed tests and the fractal dimension: as the latter increases, the former also increases. Indeed, for the two current ranges the correlation coefficients between the percentage of passed tests and $D_2$, obtained with the formula $\rho_{X,Y}=\frac{\mathbb{E}[XY]-\mathbb{E}[X]\mathbb{E}[Y]}{\sqrt{\mathbb{E}[X^2]-(\mathbb{E}[X])^2}\sqrt{\mathbb{E}[Y^2]-(\mathbb{E}[Y])^2}}$ where $\mathbb{E}[X]$ is the expectation value of $X$, are 0.66 and 0.67 respectively. However, this correlation holds only within a given chaotic zone and no longer applies when comparing different chaotic zones. For instance, although the fractal dimension is roughly twice as large in the first chaotic zone as in the second, the percentage of passed NIST tests is not higher. This suggests that, for random number generation, the specific dynamics of each chaotic zone play a dominant role, not the typical correlation dimension. Nevertheless, within a chaotic zone characterized by similar underlying dynamics, a higher correlation dimension leads to a higher percentage of passed NIST tests.\\
\begin{figure}[t]
	\centering
	\fbox{\includegraphics[width=\linewidth]{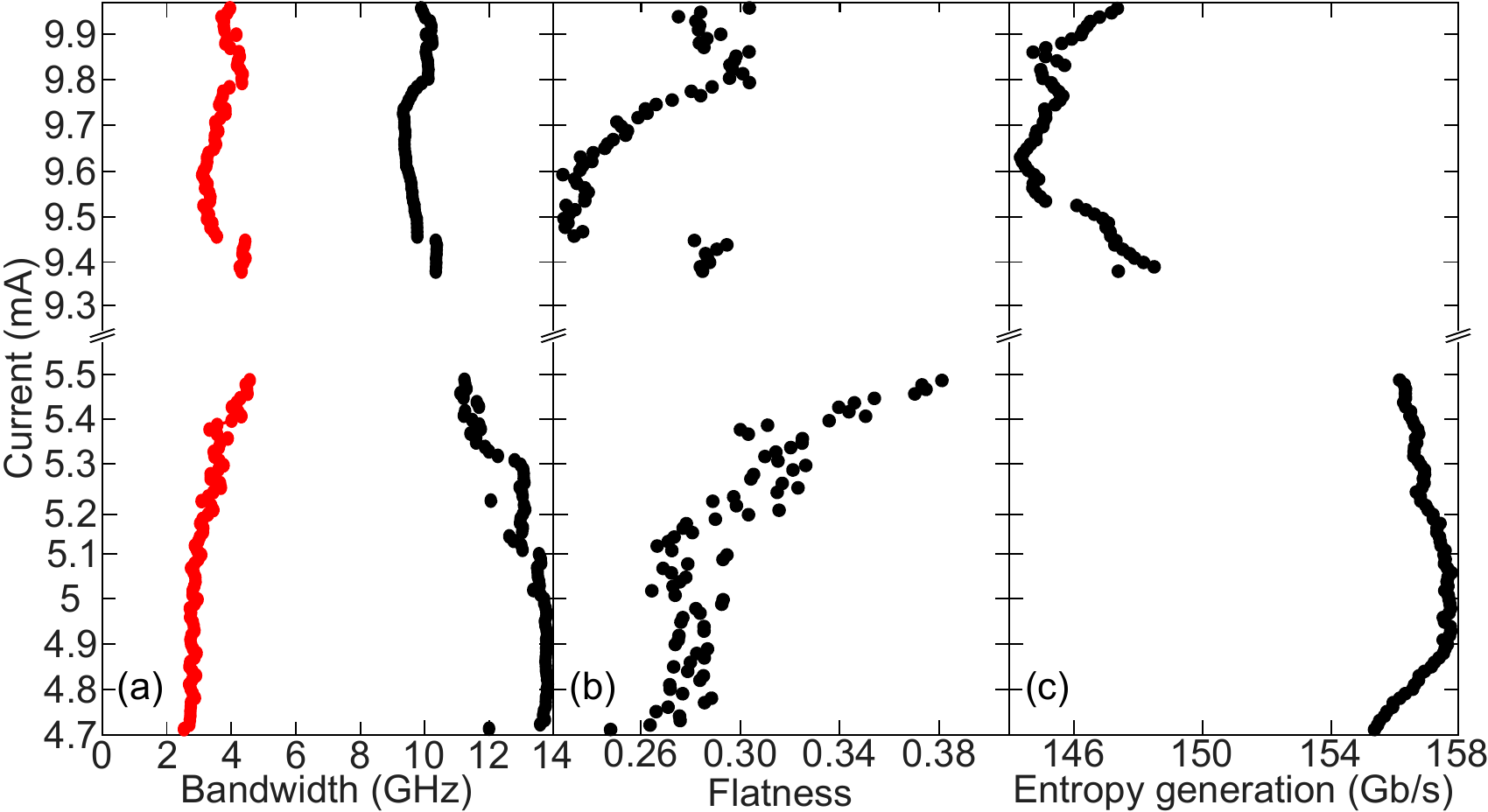}}
	\caption{(a) Evolution of the traditional (black) and effective (red) bandwidth, (b) of the spectral flatness and (c) the entropy generation for $I\in$[4.7,5.5] mA and $I\in$[9.37,9.99] mA.}
	\label{fig:fig3}
\end{figure}
\begin{figure}[t]
	\centering
	\fbox{\includegraphics[width=\linewidth]{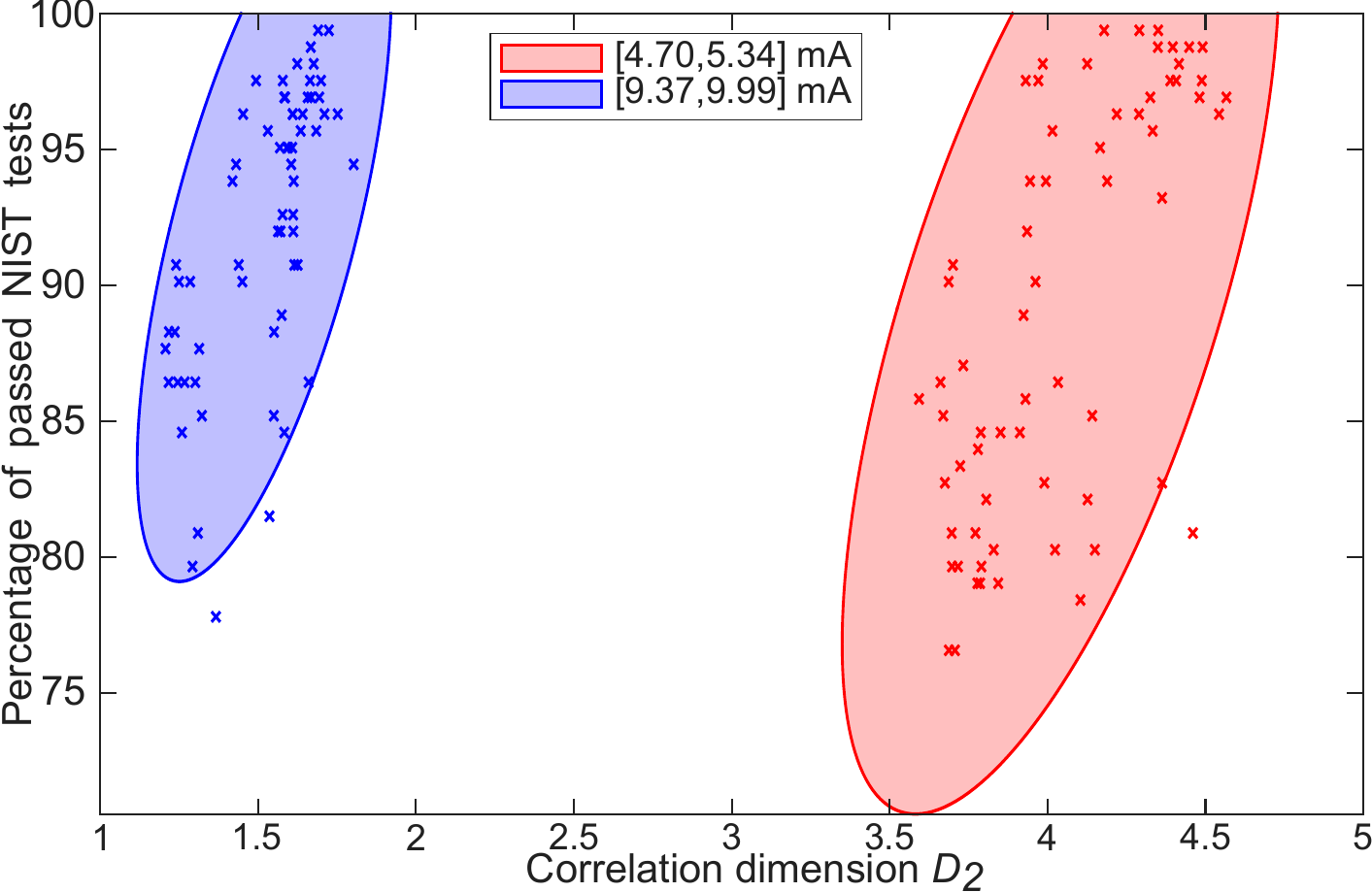}}
	\caption{Comparison between the correlation dimensions and the percentage of passed NIST tests for $I\in$[4.7, 5.34] mA (red) and $I\in$[9.37, 9.99] mA (blue). The bit streams are $2\times10^4$ bits long and are generated with a sampling rate of 1 Gb/s, thresholding with the mean voltage and a 40 ns XOR delay.}
	\label{fig:fig4}
\end{figure}
\begin{figure*}[t]
	\centering
	\fbox{\includegraphics[width=\linewidth]{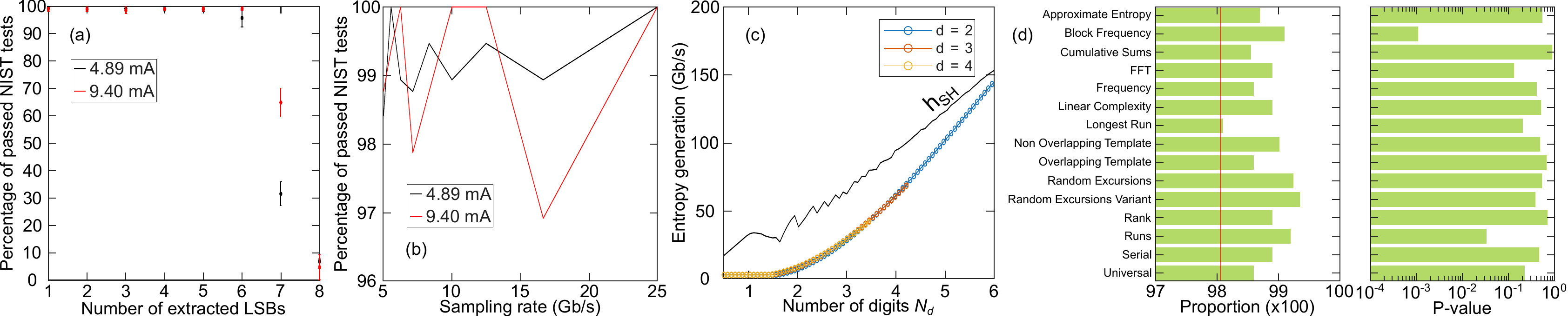}}
	\caption{(a) Evolution of the percentage of passed NIST tests for different numbers of extracted least significant bits (LSBs) for $I$ = 4.89~mA (black) and $I$ = 9.40~mA (red). The sampling rate is 1~Gb/s and a XOR delay of 40~ns is applied. (b) Percentage of passed NIST tests for different sampling rates. An XOR delay of 320~ps is applied for both injection currents and 5 LSBs (6 LSBs) are extracted for $I$ = 4.89~mA (9.40~mA). (c) Entropy generation for $I$ = 9.40~mA as a function of the number of digits for embedding dimensions 2 to 4 and the Shannon-Hartley limit (black). (d) Representative NIST SP 800-22 results for $I$ = 9.40~mA, a sampling rate of 25~Gb/s, 6 LSBs and a XOR delay of 360~ps.}
	\label{fig:fig5}
\end{figure*}
Furthermore, we investigate the performances of the RNG of the BA-VCSEL more thoroughly for two specific currents. Figures~\ref{fig:fig5}(a) and \ref{fig:fig5}(b) show the percentages of passed NIST tests for two injection currents with different numbers of extracted least significant bits (LSB) and different sampling rates, respectively. The two chosen injection currents correspond to one of the best cases (9.40~mA) and one of the worst cases (4.89~mA) in terms of passed NIST tests in Fig.~\ref{fig:fig4}. In Fig.~\ref{fig:fig5}(a), a sampling rate of 1~Gb/s and a XOR delay of 5~ns is used. This figure plots the evolution of the percentage of passed NIST tests for different numbers of extracted LSBs. For each data point, the mean value and the standard deviation are calculated on a sample of 15 series of 1 million bits. We see that $I$ = 9.40~mA allows for $\approx$100\% of passed tests at 6 LSBs while it decreases by a few percent for 4.89~mA. This indicates that for the latter injection current, 5 LSBs is the maximum number of extracted bits to pass all NIST tests. \\
\indent
Similarly Fig.~\ref{fig:fig5}(b) plots the evolution of the percentage of passed NIST tests for different sampling rates. For each data point, 1 million bits and a XOR delay of 360 ps were used. Also, for $I$ = 4.89~mA (9.40~mA), 5 LSBs (6 LSBs) were extracted. Hence, we observe that with the chosen post-processing, both injection currents pass all NIST tests for a sampling rate of 25~Gb/s. It is worth noting that we limit ourselves to a sampling rate of 25~Gb/s because the BWs of our signals lie between 10 and 15~GHz, corresponding to a sampling rate of approximately 20 to 30~Gb/s according to Nyquist criterion. With more sophisticated post-processing, it could be possible to reach a sampling rate of 50~Gb/s. However, the randomness of our bit streams might be insufficient \cite{hart2017recommendations}.\\
\indent
Figure~\ref{fig:fig5}(c) plots the evolution of the estimated entropy generation rate with respect to the number of digits for $I = 9.40$~mA and for three different embedding dimensions. The theoretical Shannon-Hartley limit $h_{\text{SH}}$ is also plotted in black \cite{hart2017recommendations}, which is proportional to the BW. As it is not quite clear which exact value to use because of the uneven shape of the RF-spectrum and its variation with the injection current, a BW of 16~GHz has been chosen as an upper limit since it is close to the values found in Fig.~\ref{fig:fig3} and because most of the RF-signal is included between 0 and 16~GHz. As expected, both entropy rates increase linearly as the number of digits increases for $N_{d} > 4$ \cite{kim2021massively,hart2017recommendations}. Also, we note that for a number of digits around 6, the estimated entropy generation is close to $h_{\text{SH}}$ indicating that for 6 LSBs the maximal possible rate is almost achieved. Hence, for 6 LSBs, an extraction of ~150~Gb/s can be expected. This matches our experimental extraction of 150~Gb/s = 6$\times$25~Gb/s for 9.40~mA. Also, while this result is lower than recent random number extractions from BA-VCSELs \cite{lu2024parallel}, we note that we are limited by the maximum entropy extraction. A more sophisticated post-processing (such as an extended XOR) may enable faster RNG but would likely compromise the randomness of the bit stream.\\
\indent
Lastly, to assess the randomness of our chaotic temporal trace at 9.40~mA, we use the NIST SP 800-22 statistical test suite \cite{sonmez2016recommendation} with a sampling rate of 25~Gb/s, 6 LSBs and a XOR delay of 360~ps. Figure~\ref{fig:fig5}(d) shows the results for 1000 series of 1 million bits, passing all tests. It is worth noting that $I = 4.89$~mA, one of the worst cases in Fig.\ref{fig:fig4}, still passes all 15 NIST tests of the SP 800-22 statistical test suite with optimized post-processing parameters, that is a sampling rate of 25~Gb/s, 5 LSBs and a XOR delay of 360~ps (see Appendix \ref{AppendixWorstCase}).
\section*{Conclusions}
\indent
In summary, we have demonstrated that BA-VCSELs show a large variety of chaotic dynamics in a large range of currents, which differ by their correlation dimensions, BW and spectral flatness. The changes of correlation dimension occur when bifurcations take place when varying the injection current and these are well identified by qualitative changes of the frequency components in the RF spectrum. Interestingly, large fractal dimension and chaos with large BW is obtained for several ranges of injection current values, enabling large entropy generation rates up to 150~Gb/s from a single VCSEL without the need for optical feedback, optical injection or even external modulation. This performance is validated by the NIST tests and is encouraging for using a BA-VCSEL as a fast, compact and stable random number generator. While faster random bit rates have been achieved with other setups, the BA-VCSELs have the great advantage of being stand-alone devices without the need for alignment-sensitive external cavities. Their simplicity and insensitivity to small pump current variations makes BA-VCSELs very promising for real-world applications that need to be compact, cost-effective and robust to environmental fluctuations. \\
\indent
We also conducted similar analyses on two other BA-VCSELs and obtained comparable results, indicating that the properties we observed are common to this model of BA-VCSELs. These findings open the possibility of using multiple VCSELs in an array to achieve extremely fast random number generation, potentially reaching several THz/s with a dozen devices. Moreover, BA-VCSELs with more transverse modes could be considered, as this may induce more bifurcations leading to a greater range of dynamics.\\
\begin{acknowledgments}
The Chair in Photonics is supported by Region Grand Est, GDI Simulation, Département de la Moselle, European Regional Development Fund, CentraleSupelec, Fondation CentraleSupelec and Eurometropole de Metz.
\end{acknowledgments}

\appendix

\section{Noise titration of chaos \label{AppendixNoiseTitration}}
\begin{figure}[!b]
	\centering
	\fbox{\includegraphics[width=0.9\linewidth]{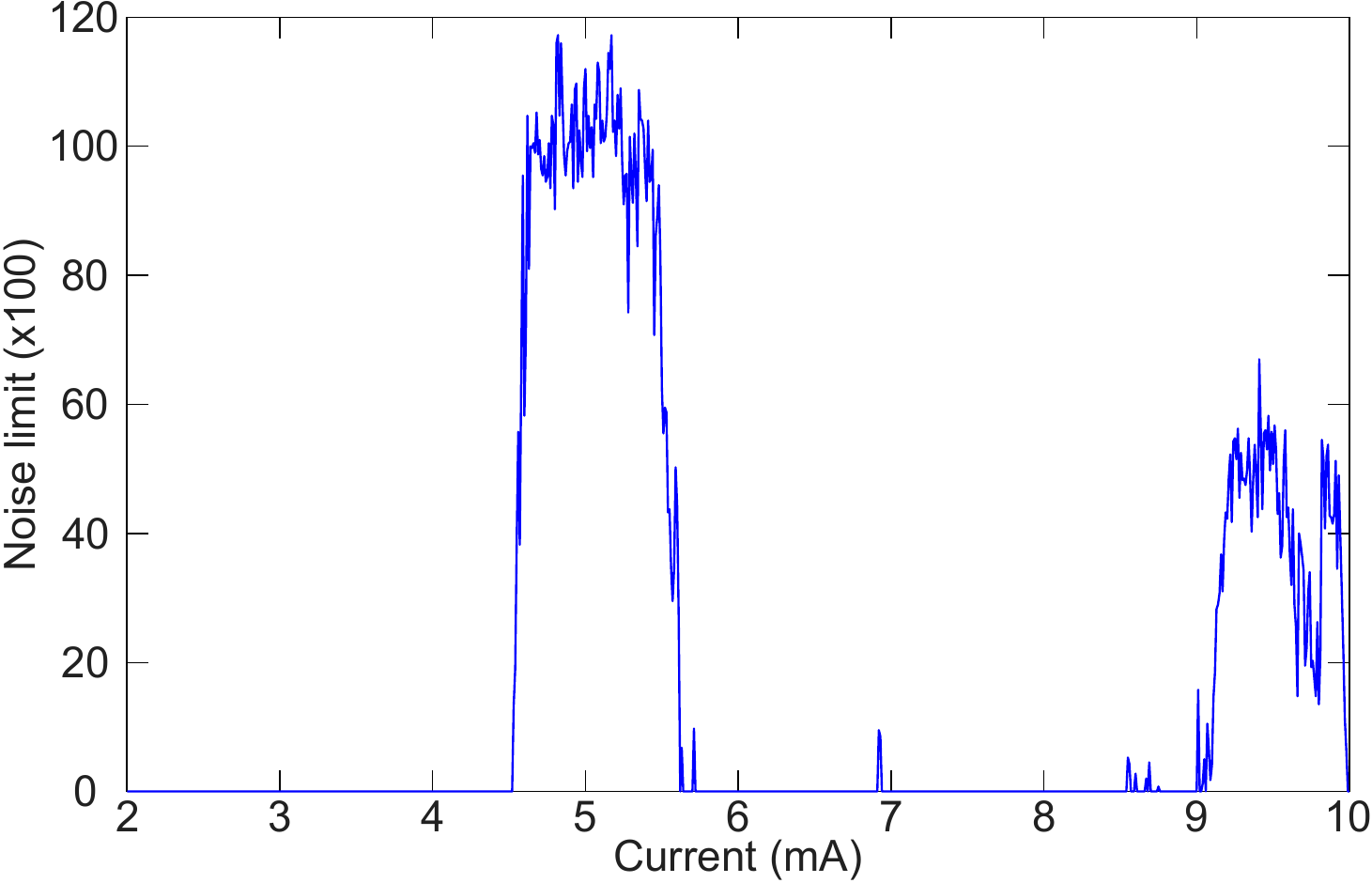}}
	\caption{Evolution of the noise limit from the noise titration method for different currents and for a depth $k$ = 10. A strictly positive noise limit indicates the presence of nonlinear dynamics.}
	\label{fig:figS1}
\end{figure}
First, we detect the current regions showing nonlinear or chaotic dynamics by means of a noise titration method \cite{Poon2001a}, which compares the prediction performance of two autoregressive models: a linear one and a nonlinear one (typically of Volterra--Wiener type). Both models are fitted to the same input time series using a memory depth $k$ to adjust their respective parameters. The core idea is to assess whether the time series evolution can be better described by a linear or by a nonlinear model.\\
For a time series $x_t$ where $t$ is an index corresponding to time, a linear autoregressive model of depth $k$ predicts the next data point via the model
\begin{equation}
	\hat{x}_t^{(\mathrm{L})} = \sum_{i=1}^{k} a_i x_{t-i} + d
	\label{linModel}
\end{equation}
where $a_i$ are the linear coefficients and $d$ is a bias term. The nonlinear model extends this by including quadratic terms to capture nonlinear interactions,
\begin{equation}
	\hat{x}_t^{(\mathrm{NL})} = \sum_{i=1}^{k} a_i x_{t-i} + \sum_{i=1}^{k} b_i x_{t-i}^2 + \sum_{1 \le i < j \le k} c_{ij} x_{t-i} x_{t-j} + d \, ,
\end{equation}
where $b_i$ and $c_{ij}$ are additional coefficients estimated via least-squares regression. The prediction errors $E_{L}$ and $E_{NL}$ are then computed as the mean squared error (MSE) over the time series
\begin{equation}
	E_\mathrm{L} = \frac{1}{N} \sum_{t=1}^{N} \left(x_t - \hat{x}_t^{(\mathrm{L})} \right)^2, \quad
	E_\mathrm{NL} = \frac{1}{N} \sum_{t=1}^{N} \left(x_t - \hat{x}_t^{(\mathrm{NL})} \right)^2 \, .
\end{equation}

In the presence of nonlinear dynamics, the nonlinear model typically achieves a lower MSE than the linear model. To quantify the robustness of the nonlinear behavior, we progressively add zero-mean Gaussian noise $\eta(t)$ of variance $\sigma^2 = \alpha\text{Var}[x(t)]$ to the original time series:
\begin{equation}
	x_n(t) = x(t) + \eta(t)
\end{equation}
\indent
For each noise level $\alpha$, both models are refitted and prediction errors recomputed. The noise limit (NL) is defined as the smallest noise level $\alpha_\mathrm{NL}$ at which the linear model begins to outperform the nonlinear model. Hence, a low NL (close to 0\%) indicates that a linear or highly noisy dynamics is more likely, while a high NL strongly indicates a nonlinear dynamic. By performing this procedure for different injection currents, one can systematically map the operating regions in which nonlinear behavior dominates the laser dynamics as plotted in Fig.~\ref{fig:figS1} for a depth $k$ = 10.

\section{Implementation of the Grassberger--Procaccia Algorithm \label{GPAlgo}}
\indent

For our noisy experimental time series, it was necessary to reduce the noise prior to further analysis. To achieve this, we applied a Singular Spectrum Analysis (SSA) algorithm \cite{golyandina2001analysis}. SSA consists of a phase-space reconstruction through time-delay embedding, where each embedded vector is defined as 
\begin{equation}
	\mathbf{X}_i = \left[ x(t_i), x(t_i+1), x(t_i+2), \dots, x(t_i+(m-1)) \right]
	\label{SSAPhaseSpace}
\end{equation}
with $m$ a relatively large embedding dimension (in our case, $m \in [40,80]$) to allow the separation of the main signal components from noise and a unitary embedding delay. In our case $x(t)$ is the signal of the photodetector measured by the oscilloscope. \\
Next, the trajectory matrix $\mathbf{X}$, defined by stacking the embedded vectors as
\begin{equation}
	\mathbf{X} =
	\begin{bmatrix}
		\mathbf{X}_1 \\
		\mathbf{X}_2 \\
		\vdots \\
		\mathbf{X}_N
	\end{bmatrix},
\end{equation}
is decomposed using singular value decomposition (SVD) such that $\mathbf{X} = \mathbf{U} \mathbf{\Sigma} \mathbf{V}^\top$, where $\mathbf{\Sigma} = \mathrm{diag}(\sigma_1, \sigma_2, \dots, \sigma_m)$ contains the singular values in decreasing order. By retaining only the first $k$ singular values corresponding to the most prominent components, the principal dynamics of the signal is preserved while the noisy part is removed.\\
Finally, the denoised time series is reconstructed from the truncated SVD as $\hat{\mathbf{X}} = \mathbf{U}_k \mathbf{\Sigma}_k \mathbf{V}_k^\top$ where $\mathbf{U}_k$, $\mathbf{\Sigma}_k$, and $\mathbf{V}_k$ contain only the selected $k$ components. It should be noted that keeping too few singular values reduces the signal complexity towards that of a periodic signal, whereas keeping too many fails to sufficiently remove noise. Here, we chose $k$ as the index preceding the onset of a plateau or a gradual decay in the singular values ordered in decreasing magnitude \cite{golyandina2001analysis}.\\
\indent
Afterwards, the correlation dimension was estimated using the Grassberger--Procaccia (GP) algorithm \cite{grassberger1983measuring} applied to our denoised time series after subtracting the mean and normalizing the signal $x(t)$.\\
\indent
To apply the GP algorithm, we first extract a certain number of data points from the full time series, in our case 400$\times10^3$. Then we reconstruct the phase space from the extracted time series using a time-delay embedding, where the vectors are defined as
\begin{equation}
	\mathbf{X}_i = \left[ x(t_i),\, x(t_i+\tau),\, x(t_i+2\tau),\, \dots,\, x(t_i+(m-1)\tau) \right]
	\label{phaseSpaceVector}
\end{equation}
with $\tau$ the embedding delay and $m$ the embedding dimension. The delay $\tau$ was selected as the first minimum of the mutual information of the time series $x$. Next, $m$ was progressively increased until a convergence of the estimated fractal dimension was observed.\\
To estimate this dimension, its estimator, the correlation integral
\begin{equation}
	C_m(r) = \frac{2}{N(N-1)}\sum_{\substack{i<j \\ |i-j| > T}} H\!\left(r - \lVert \mathbf{X}_i - \mathbf{X}_j \rVert \right)
	\label{CorrelationIntegral}
\end{equation}
can be computed, where $H(\cdot)$ is the Heaviside step function, $r$ is the phase-space radius, $N$ is the number of embedded vectors and $T$ is the Theiler window. The Theiler window excludes temporally close points, reducing short-term correlations that could bias the scaling behavior $C_m(r) \propto r^{D_2}$ used to estimate the correlation dimension for chaotic systems \cite{theiler1986spurious}. This Theiler window is determined through a space-time separation plot, which depicts the probability $P(|s_i - s_j| < \epsilon | |i - j| = \Delta t)$ that a pair of points is separated by a spatial distance $\epsilon$ for a given temporal lag $\Delta t$. The Theiler window $T$ is identified as the time lag beyond which the observed contour lines become independant of the time lag \cite{kantz2003nonlinear}. In our case, we chose $T$ = 250~ns. The correlation dimension $D_2$ is finally obtained from the slope 
\begin{equation}
	D_2 = \lim_{r \to 0} \frac{d \log C_m(r)}{d \log r}
	\label{CorrelationDimension}
\end{equation}
in the scaling region. 

In practice, $D_2$ was estimated from the slope of the linear scaling region in a $\log[C_m(r)]$ versus $\log(r)$ representation [see Fig.~1(d) in the main text]. Particular care was taken in selecting this scaling region to avoid bias from measurement noise at small radii and finite-size effects at large radii. The robustness of the results was verified by varying the embedding delay, embedding dimension, the length of the analyzed time series and the number of extracted points from the full time series.\\

\section{Entropy rate estimation \label{EGEstimation}}
\indent

To estimate the entropy generation rate of the measured time series, we employed the Cohen--Procaccia algorithm \cite{hart2017recommendations,kim2021massively,cohen1985computing}. This method evaluates the entropy rate by quantifying how the number of distinguishable trajectories grows in the reconstructed phase space as the resolution is refined.\\
\indent
First, the time series $x(t)$ is quantized by defining a bin size 
\begin{equation}
	\epsilon = \frac{x_{\max}-x_{\min}}{2^{N_{\text{d}}}}
\end{equation}
determined from the dynamic range of the signal, where $x_{\max}$ and $x_{\min}$ are the maximum and minimum intensity values of the temporal trace, and $N_{\text{d}}$ is the number of quantization digits used for the analysis.\\
The time series is embedded into a $d$-dimensional phase space using a unitary embedding delay ($\tau = 1$), with embedded vectors
\begin{equation}
	\mathbf{X}_i = [x(t_i), x(t_i+1), x(t_i+2), \dots, x(t_i+(d-1))] \, .
\end{equation}
This choice for the embedding delay ensures that all available information is preserved, as any larger interval would result in undersampling and a subsequent loss of extractable entropy. Also, the embedding dimensions are chosen to ensure that the entropy estimates for different dimensions remain consistent. This indicates that the calculated entropy rate effectively captures the unpredictable content of the signal and is not biased by the specific choice of embedding parameters. \\
\indent Then, among all reconstructed vectors $\mathbf{X}_i$, a subset of $N$ reference vectors is selected (typically $N=3\times10^4$). For each reference vector $\mathbf{X}_k$, the fraction $f_k(\epsilon)$ of other neighboring vectors located within a $d$-dimensional hypercube of side length $\epsilon$ centered on $\mathbf{X}_k$ is computed. The $d$-dimensional pattern entropy is estimated as 
\begin{equation}
	H_d(\epsilon) =
	-\frac{1}{N}
	\sum_{k=1}^{N}
	\log_2\!\left[f_k(\epsilon)\right]  \, .
	\label{PatternEntropy}
\end{equation}
At last, the entropy rate can be obtained from the incremental growth of the pattern entropy with the embedding dimension,
\begin{equation}
	h_{\mathrm{CP}}(\epsilon,d)
	=
	\frac{1}{\tau_{s}}
	\left[
	H_d(\epsilon) - H_{d-1}(\epsilon)
	\right] \, ,
\end{equation}
where $\tau_{s}^{-1}$ is the sampling rate. This quantity provides an estimate of the entropy production rate of the dynamical system. A higher value of $h_{\mathrm{CP}}$ indicates a larger degree of dynamical complexity and unpredictability in the time series.\\
\indent
We also computed the information-theoretical limit derived from the Shannon--Hartley theorem. This limit corresponds to the maximum entropy rate that can be extracted from the measured signal. In the context of digitized chaotic signals, this limit can be written as:

\begin{equation}
	h_{SH} =
	\min\!\left(\frac{1}{\tau_{s}},\, 2f_{\mathrm{BW}}\right)
	\left[
	N_{\mathrm{d}}
	-
	D_{\mathrm{KL}}\!\left[p(x)\|u(x)\right]
	\right],
\end{equation}
where $\tau_{s}^{-1}$ is the sampling rate, $f_{\mathrm{BW}}$ denotes the effective bandwidth of the signal and $N_{\mathrm{d}}$ is the number of digitization bits. The prefactor $\min\!\left(\frac{1}{\tau_{s}},\, 2f_{\mathrm{BW}}\right)$ reflects the maximum rate at which statistically independent samples can be obtained, which is limited either by the sampling rate $\frac{1}{\tau_{s}}$ or by the Nyquist rate $2f_{\mathrm{BW}}$ associated with the signal bandwidth.\\
The term $D_{\mathrm{KL}}[p(x)\|u(x)]$ is the Kullback–Leibler divergence between the measured intensity probability density function $p(x)$ and the uniform distribution $u(x)$ defined over the same digitization range, 
\begin{equation}
	D_{\mathrm{KL}}\!\left[p(x)\|u(x)\right]
	=
	\sum_x
	p(x)
	\log_2
	\left(
	\frac{p(x)}{u(x)}
	\right) \, .
\end{equation}
This divergence quantifies the deviation of the measured intensity distribution from an ideal uniform distribution. A perfectly uniform distribution would yield $D_{\mathrm{KL}}=0$, allowing the full $N_{\mathrm{d}}$ bits of entropy per sample to be extracted. In practice, non-uniformities in the signal distribution reduce the achievable entropy rate. The quantity $h_{SH}$ therefore represents the theoretical upper bound for the entropy generation rate of the digitized signal.

\section{NIST SP 800-22 tests results for the worst case of chaos \label{AppendixWorstCase}}
\indent

\begin{figure}[ht]
	\centering
	\fbox{\includegraphics[width=\linewidth]{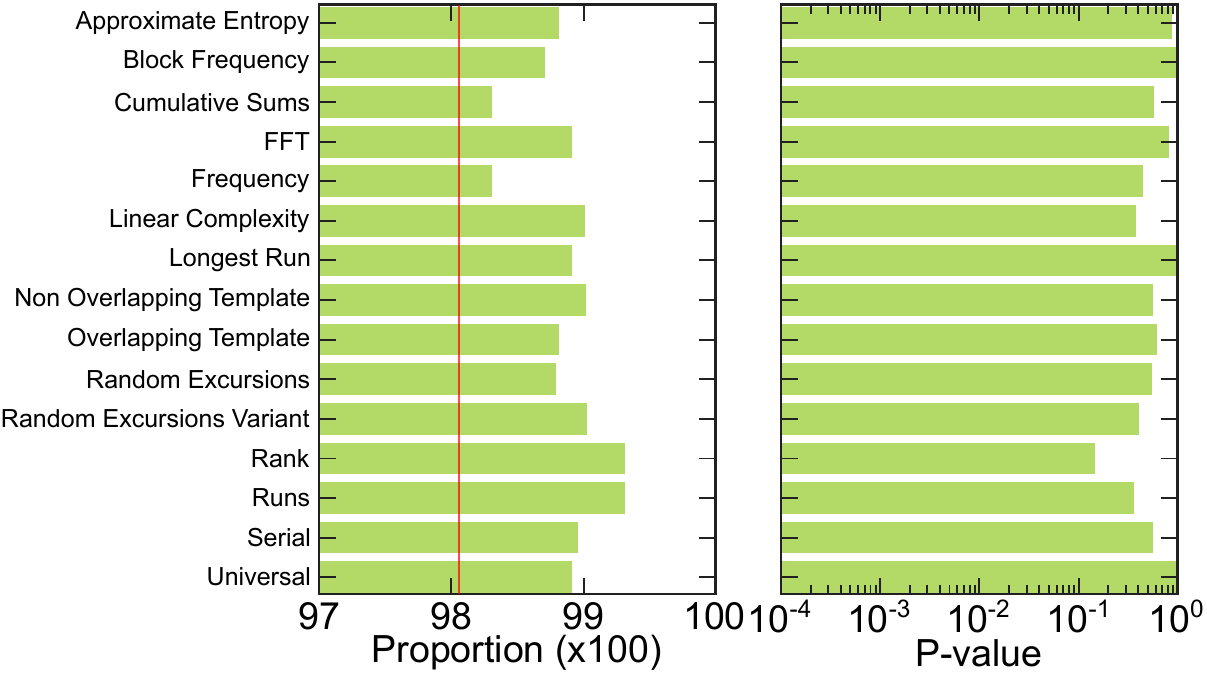}}
	\caption{Representative NIST SP 800-22 results for $I$ = 4.89~mA, a sampling rate of 25~Gb/s, 5 LSBs and a XOR delay of 360~ps.}
	\label{fig:figS2}
\end{figure}

To ensure that even in the case ($I$ = 4.89~mA) our BA-VCSEL can still act as a high quality fast random numbers generator, we used the NIST SP 800-22 test suite with a sampling rate of 25~Gb/s, a XOR delay of 360~ps and 5 extracted LSBs in this case. Figure~\ref{fig:figS2} shows the results for 1000 series of 1 million bits. Hence, even in this worst case of Fig.~4, our BA-VCSEL can still act as a physical random number generator with an experimental extraction of 5$\times$25~Gb/s = 125~Gb/s.

\nocite{*}
\bibliography{aapmsamp}

@article{soriano2013complex,
	title={Complex photonics: Dynamics and applications of delay-coupled semiconductors lasers},
	author={Soriano, Miguel C and Garc{\'\i}a-Ojalvo, Jordi and Mirasso, Claudio R and Fischer, Ingo},
	journal={Reviews of Modern Physics},
	volume={85},
	number={1},
	pages={421--470},
	year={2013},
	publisher={APS}
}

@article{lang1980external,
	title={External optical feedback effects on semiconductor injection laser properties},
	author={Lang, Roy and Kobayashi, Kohroh},
	journal={IEEE Journal of Quantum Electronics},
	volume={16},
	number={3},
	pages={347--355},
	year={1980},
	publisher={IEEE}
}

@article{bouchez2021optimized,
	title={Optimized properties of chaos from a laser diode},
	author={Bouchez, Guillaume and Malica, Tushar and Wolfersberger, Delphine and Sciamanna, Marc},
	journal={Physical Review E},
	volume={103},
	number={4},
	pages={042207},
	year={2021},
	publisher={APS}
}

@article{kovanis1995instabilities,
	title={Instabilities and chaos in optically injected semiconductor lasers},
	author={Kovanis, Vassilios and Gavrielides, Athanasios and Simpson, TB and Liu, Jia-Ming},
	journal={Applied Physics Letters},
	volume={67},
	number={19},
	pages={2780--2782},
	year={1995},
	publisher={American Institute of Physics}
}

@article{mercadier2024optical,
	title={Optical chaos synchronization in a cascaded injection experiment},
	author={Mercadier, Jules and Doumbia, Yaya and Bittner, Stefan and Sciamanna, Marc},
	journal={Optics Letters},
	volume={49},
	number={10},
	pages={2613--2616},
	year={2024},
	publisher={Optica Publishing Group}
}

@article{virte2013deterministic,
	title={Deterministic polarization chaos from a laser diode},
	author={Virte, Martin and Panajotov, Krassimir and Thienpont, Hugo and Sciamanna, Marc},
	journal={Nature Photonics},
	volume={7},
	number={1},
	pages={60--65},
	year={2013},
	publisher={Nature Publishing Group UK London}
}

@article{virte2013bifurcation,
	title={Bifurcation to nonlinear polarization dynamics and chaos in vertical-cavity surface-emitting lasers},
	author={Virte, Martin and Panajotov, Krassimir and Sciamanna, Marc},
	journal={Physical Review A},
	volume={87},
	number={1},
	pages={13834},
	year={2013},
	publisher={APS}
}

@article{mercadier2025chaos,
	title={Chaos from a free-running broad-area VCSEL},
	author={Mercadier, Jules and Bittner, Stefan and Rontani, Damien and Sciamanna, Marc},
	journal={Optics Letters},
	volume={50},
	number={3},
	pages={796--799},
	year={2025},
	publisher={Optica Publishing Group}
}

@article{sanvert2025polarization,
	title={Polarization and transverse-mode nonlinear dynamics in a multimode VCSEL},
	author={Sanvert, Yohann G and Mercadier, Jules and Bittner, Stefan and Valle, Angel and Sciamanna, Marc},
	journal={Optics Letters},
	volume={50},
	number={24},
	pages={7645--7648},
	year={2025},
	publisher={Optica Publishing Group}
}

@article{bittner2022complex,
	title={Complex nonlinear dynamics of polarization and transverse modes in a broad-area VCSEL},
	author={Bittner, Stefan and Sciamanna, Marc},
	journal={APL Photonics},
	volume={7},
	pages={126108},
	number={12},
	year={2022},
	publisher={AIP Publishing}
}

@article{lu2024parallel,
	title={Parallel on-chip physical random number generator based on self-chaotic dynamics of free-running broad-area VCSEL array},
	author={Lu, Hang and Alkhazragi, Omar and Wang, Yue and Ng, Tien Khee and Ooi, Boon S},
	journal={IEEE Journal of Selected Topics in Quantum Electronics},
	volume={31},
	number={2: Pwr. and Effic. Scaling in Semiconductor Lasers},
	pages={1--11},
	year={2024},
	publisher={IEEE}
}

@article{hirano2009characteristics,
	title={Characteristics of fast physical random bit generation using chaotic semiconductor lasers},
	author={Hirano, Kunihito and Amano, Kazuya and Uchida, Atsushi and Naito, Sunao and Inoue, Masaki and Yoshimori, Shigeru and Yoshimura, Kazuyuki and Davis, Peter},
	journal={IEEE Journal of Quantum Electronics},
	volume={45},
	number={11},
	pages={1367--1379},
	year={2009},
	publisher={IEEE}
}

@article{tseng2025scalable,
	title={Scalable ultrafast random bit generation using wideband chaos-based entropy sources},
	author={Tseng, Chin-Hao and Uchida, Atsushi and Hwang, Sheng-Kwang},
	journal={arXiv:2512.24716},
	year={2025}
}

@article{virte2014physical,
	title={Physical random bit generation from chaotic solitary laser diode},
	author={Virte, Martin and Mercier, Emeric and Thienpont, Hugo and Panajotov, Krassimir and Sciamanna, Marc},
	journal={Optics Express},
	volume={22},
	number={14},
	pages={17271--17280},
	year={2014},
	publisher={Optical Society of America}
}

@article{li2026solitary,
	title={Solitary Chaotic Semiconductor Laser Diode for Parallel Random Number Generation and Optical Decision Making},
	author={Li, Jiacheng and Wang, Fei and Deng, Tao},
	journal={Journal of Lightwave Technology},
	volume={44},
	number={4},
	pages={1420--1429},
	year={2026},
	publisher={IEEE}
}

@article{uchida2008fast,
	title={Fast physical random bit generation with chaotic semiconductor lasers},
	author={Uchida, Atsushi and Amano, Kazuya and Inoue, Masaki and Hirano, Kunihito and Naito, Sunao and Someya, Hiroyuki and Oowada, Isao and Kurashige, Takayuki and Shiki, Masaru and Yoshimori, Shigeru and others},
	journal={Nature Photonics},
	volume={2},
	number={12},
	pages={728--732},
	year={2008},
	publisher={Nature Publishing Group UK London}
}

@article{mercadier2026synchronization,
	title={Synchronization of complex spatio-temporal dynamics with lasers},
	author={Mercadier, Jules and Bittner, Stefan and Sciamanna, Marc},
	journal={Light: Science \& Applications},
	volume={15},
	number={1},
	pages={131},
	year={2026},
	publisher={Nature Publishing Group UK London}
}

@article{grassberger1983measuring,
	title={Measuring the strangeness of strange attractors},
	author={Grassberger, Peter and Procaccia, Itamar},
	journal={Physica D: nonlinear phenomena},
	volume={9},
	number={1-2},
	pages={189--208},
	year={1983},
	publisher={Elsevier}
}

@article{lin2012effective,
	title={Effective bandwidths of broadband chaotic signals},
	author={Lin, Fan-Yi and Chao, Yuh-Kwei and Wu, Tsung-Chieh},
	journal={IEEE Journal of Quantum Electronics},
	volume={48},
	number={8},
	pages={1010--1014},
	year={2012},
	publisher={IEEE}
}

@article{johnston2002transform,
	title={Transform coding of audio signals using perceptual noise criteria},
	author={Johnston, James D},
	journal={IEEE Journal on Selected Areas in Communications},
	volume={6},
	number={2},
	pages={314--323},
	year={2002},
	publisher={IEEE}
}

@article{cohen1985computing,
	title={Computing the Kolmogorov entropy from time signals of dissipative and conservative dynamical systems},
	author={Cohen, Aviad and Procaccia, Itamar},
	journal={Physical review A},
	volume={31},
	number={3},
	pages={1872},
	year={1985},
	publisher={APS}
}

@article{kim2021massively,
	title={Massively parallel ultrafast random bit generation with a chip-scale laser},
	author={Kim, Kyungduk and Bittner, Stefan and Zeng, Yongquan and Guazzotti, Stefano and Hess, Ortwin and Wang, Qi Jie and Cao, Hui},
	journal={Science},
	volume={371},
	number={6532},
	pages={948--952},
	year={2021},
	publisher={American Association for the Advancement of Science}
}

@article{theiler1986spurious,
	title={Spurious dimension from correlation algorithms applied to limited time-series data},
	author={Theiler, James},
	journal={Physical Review A},
	volume={34},
	number={3},
	pages={2427},
	year={1986},
	publisher={APS}
}

@article{hart2017recommendations,
	title={Recommendations and illustrations for the evaluation of photonic random number generators},
	author={Hart, Joseph D and Terashima, Yuta and Uchida, Atsushi and Baumgartner, Gerald B and Murphy, Thomas E and Roy, Rajarshi},
	journal={APL Photonics},
	volume={2},
	pages={090901},
	number={9},
	year={2017},
	publisher={AIP Publishing}
}

@techreport{sonmez2016recommendation,
	title={Recommendation for the entropy sources used for random bit generation},
	author={S{\"o}nmez Turan, Meltem and Barker, Elaine and Kelsey, John and McKay, Kerry and Baish, Mary and Boyle, Michael},
	year={2016},
	institution={National Institute of Standards and Technology}
}

@book{golyandina2001analysis,
	title={Analysis of time series structure: SSA and related techniques},
	author={Golyandina, Nina and Nekrutkin, Vladimir and Zhigljavsky, Anatoly A},
	year={2001},
	publisher={CRC press}
}

@article{adachihara1993spatiotemporal,
	title={Spatiotemporal chaos in broad-area semiconductor lasers},
	author={Adachihara, H and Hess, O and Abraham, E and Ru, P and Moloney, JV},
	journal={Journal of the Optical Society of America B},
	volume={10},
	number={4},
	pages={658--665},
	year={1993},
	publisher={Optical Society of America}
}

@Article{Sciamanna2015,
  author    = {Sciamanna, M. and Shore, K. A.},
  journal   = {Nature Photonics},
  title     = {Physics and applications of laser diode chaos},
  year      = {2015},
  pages     = {151},
  volume    = {9},
  doi       = {10.1038/nphoton.2014.326}
}

@Article{Poon2001a,
  author    = {Chi-Sang Poon and Mauricio Barahona},
  journal   = {Proc. Nat. Acad. Sci.},
  title     = {Titration of chaos with added noise},
  year      = {2001},
  month     = {jun},
  number    = {13},
  pages     = {7107--7112},
  volume    = {98},
  comment   = {Introduces the noise-titration technique to detect chaotic dynamics},
  doi       = {10.1073/pnas.131173198}
}

@book{kantz2003nonlinear,
	title={Nonlinear time series analysis},
	author={Kantz, Holger and Schreiber, Thomas},
	year={2003},
	publisher={Cambridge university press}
}

\end{document}